\documentclass[]{article}
\usepackage[utf8]{inputenc}
\usepackage[utf8]{inputenc}
\usepackage{graphicx}
\usepackage{subfigure}

\usepackage{amssymb}
\usepackage{amsthm}
\usepackage{url}
\usepackage{amsmath}
\usepackage{subfigure}
\usepackage{textcomp}
\usepackage{xcolor}

\usepackage[affil-it]{authblk}

\usepackage[margin=3cm]{geometry}

\graphicspath{{figures/}}

\title{Tactile sensing and computing on a random network of conducting fluid channels}

\author{Alessandro Chiolerio}
\affil{Center for Sustainable Future Technologies,
Istituto Italiano di Tecnologia, \\ Torino, Italy}

\author{Andrew Adamatzky}
\affil{Unconventional Computing Laboratory, UWE, \\ Bristol, UK}

\begin{document}

\maketitle

\begin{abstract}
\noindent
Liquid electronic sensors are typically based on regular arrays of channels filled with a conductive liquid. We propose that a random planar network of conducting liquid allows us for a wider spectrum of electrical responses and localisation of tactile stimuli. We also speculate that a computation protocol can be implemented on such a network, featuring mechanical inputs and electrical outputs. Our results pave a way towards future developments on sensing and computing wearables with disordered sensing networks structure. 

\vspace{3mm}

\end{abstract}

\section{Introduction}

Liquid electronics~\cite{gao2012direct,liu2019reconfigurable,nathan2012flexible,dickey2017stretchable} employs conductive liquids as wires~\cite{palleau2013self}, transistors~\cite{gkoupidenis2015neuromorphic,erokhin2005hybrid}, capacitors~\cite{okuzaki2014ionic,shi2019liquid,elschner2010pedot,scott2013gallium} and memristors~\cite{xiao2016energy,chen2014polymer,battistoni2017organic,koo2011towards,sheng2017transporting,chiolerio2016}. The advantages of the liquid designs is that they can be implemented inside an insulating polymer sheet in the networks of channels and thus become stretchable and foldable~\cite{rogers2010materials,wagner2012materials,fan2014fractal,kim2012flexible,ahn2012stretchable,yang2011silver,lu2014flexible,siegel2010foldable,wang2013user}. This approach represents a fundamental step towards new paradigms of functional, active and explorative robotics~\cite{chiolerio2017smart}.

Most published designs are based on regular arrangements of liquid electronic elements. This is advantageous in terms of controllability and repeatability of the results, yet disadvantageous in terms of fault tolerance and lack of novel behaviour. For the given reason, in the present study we consider a proximity graph, the Delaunay triangulation, constructed on the random planar set. Given a planar finite set ${\bf V}$ the Delaunay triangulation~\cite{delaunay1934sphere} ${\mathcal D}({\bf V})=\langle {\bf V}, {\bf E} \rangle$ is a graph  subdividing the space onto triangles with vertices in ${\bf V}$ and edges in $\bf E$ 
where the circumcircle of any triangle contains no points of $\bf V$ other than its vertices. 
Neighbours of a node $v \in \bf V$ are nodes from $\bf V$ connected with $v$ by edges from $\bf E$. Why do we use a random Delaunay triangulation? This triangulation gives us a sufficient yet economical (in terms of edges) coverage of a random planar set thus becoming a good candidate for tactile image recognition. In fact results on fingerprint indexing based on Delaunay triangulation~\cite{bebis1999fingerprint,wang2006delaunay}, including cancellable fingerprint templates for low resolution  wearable devices~\cite{lee2019cancelable}, show the full perspective of the triangulation. 

We use PEDOT:PSS as a liquid conductor. This conducive polymer is proved to work sufficiently well in electrodes for stretchable electronics and have been proposed in \cite{li2019pedot}, 
piezoresistive pulse sensor~\cite{jang2019flexible},
PEDOT cellulose conductive paper~\cite{fu2020flexible},
sulfonated lignin based PEDOT sensor~\cite{wang2019biocompatible},
pressure sensor~\cite{chegini2019ti},
humidity sensor~\cite{kang2019real},  
PEDOT:PSS and ionic liquid composites for thermal electronics~\cite{kee2019highly}, and is considered to be future corner stone of bioelectronics~\cite{donahue2020tailoring}. Furthermore it is less sensitive to temperature variations in comparison to liquid metals, where the liquid state is lost when the environment is sufficiently cold (typically below 283 K), and a better conductor and less chemically aggressive than ionic liquids.

The paper is structured as follows. Section~\ref{experimental} introduces experimental techniques. Spatio-temporal resolution of the network is analysed in Sect.~\ref{sensing}. A path towards mechano-electrical computing with the network is outlined in Sect.~\ref{computation}.
 
\section{Experimental}
\label{experimental}

\begin{figure}[!tbp]
\centering
  \includegraphics [height=5cm] {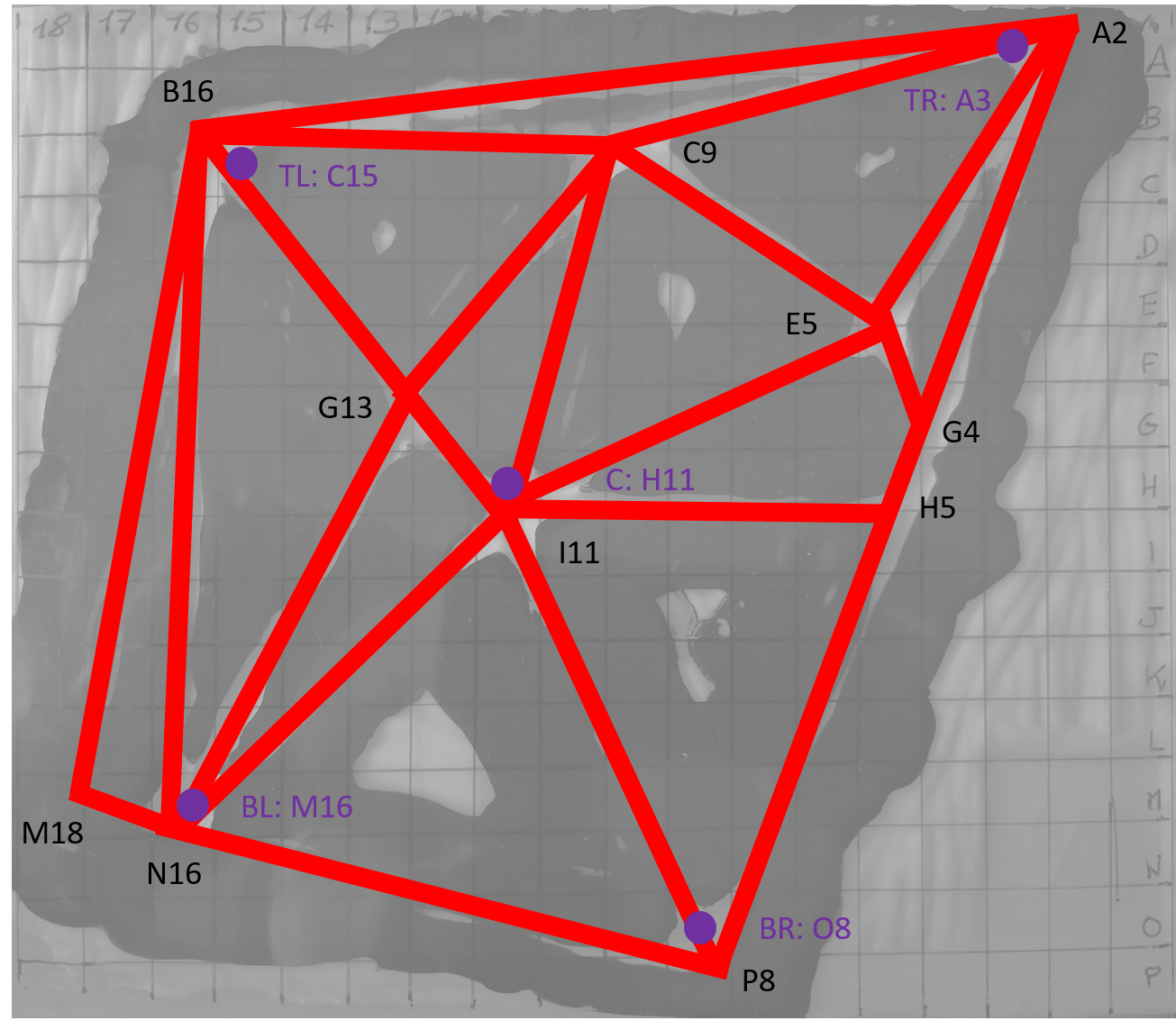}
  \caption{Sketch showing the channel network carved in the PDMS skin, with coordinates of nodes and holes for connecting electrodes. Each letter/number in the coordinate system corresponds to a square with side of 1 cm.}
  \label{fgr:E1}
\end{figure}

The flexible sensing and computing skin circuit stack is structured as follows: from bottom, a wedge-shaped polydimethylsiloxane skin (PDMS, R10PRO and R30PRO from RESCHIMICA, Via Borromini, 50 Tavarnelle Val di Pesa 50028 Firenze, Italy)  containing Delaunay triangulation channels, liquid rubber sealant (Gomma Liquida, UHU Bostik SpA Via G.B. Pirelli,19 20124 Milano, Italy), transparent rigid polyethylene terephthalate membrane (PET, 0.2 mm thickness). The skin measures 20$\times$16~cm and has a thickness ranging from 10 to 2~mm, linearly thinning from top to bottom; carved channels have a length comprised between 1.5 and 14~cm, a width of 4~mm and a depth of 2~mm. The PET membrane measures 18$\times$16 cm and is printed with a checker-board (1~cm-side squares) to support the correct stimulus application. Channels are first hand-carved into a Linoleum block (MasterCut from Essdee, Educational Art \& Craft Supplies Limited
Frederick Road, Hoo Farm Industrial Estate, Kidderminster, U.K., 300 $\times$200$\times$4~mm), then PDMS two components are mixed and poured onto the linoleum in a rectangular container. When PDMS is cured, the skin is detached from the linoleum master and liquid rubber is painted on top of the area surrounding the channels, finally PET foil with printed grid is positioned on top of the skin. Curing liquid rubber requires 1 week and some air circulation, granted by holes across PET foil done in five specific positions (see Fig.~\ref{fgr:E1}). The channel connection nodes positions are shown in black, holes across PET for filling Poly(3,4-ethylenedioxythiophene)poly(styrenesulfonate) (PEDOT:PSS) and inserting needle electrodes are shown in violet. PEDOT:PSS (CleviosTM PH 1000, Aqueous PEDOT/PSS dispersion, Heraeus Holding GmbH, Heraeusstraße 12-14, D-63450 Hanau, Germany) was injected using a lab syringe, filling the channels requires careful extraction of air bubbles, slightly facilitated by the wedge, and an overall volume of 10~ml of conductive liquid. Two needle electrodes (MN4022D10S subdermal electrodes from SPES MEDICA SRL Via Buccari 21 16153 GENOVA, Italy, 0.4~mm diameter, 22~mm in length) were inserted in holes to contact directly the conductive liquid, and connected to a Keithley 2635A multimeter (for DC characterization) and an Agilent E4980A precision LCR
meter (for AC characterization, in the range from 20 Hz up to 2 MHz).

\section{Sensing and spatio-temporal resolution}
\label{sensing}

\begin{figure}[!tbp]
\centering
\subfigure[]{\includegraphics[scale=0.27]{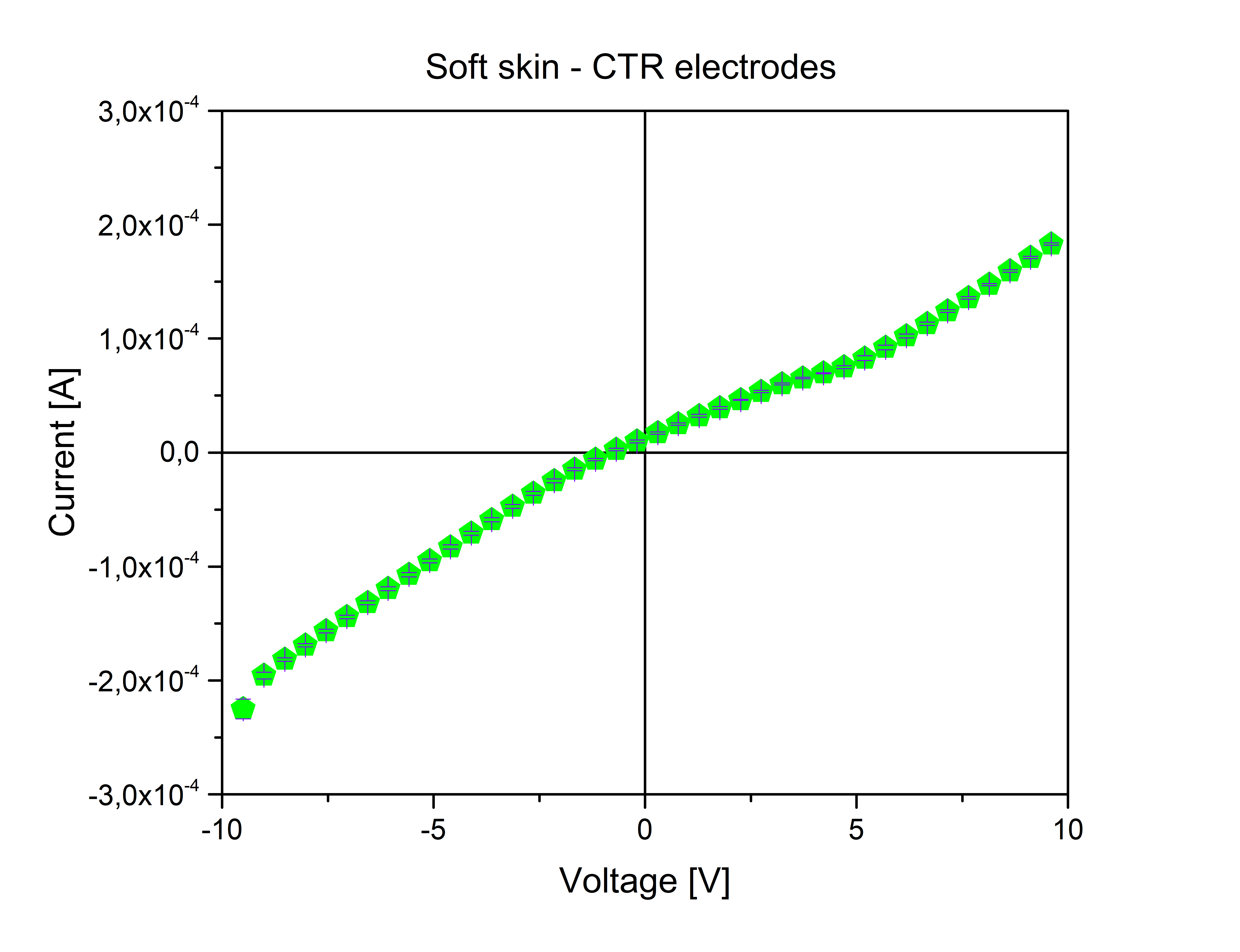}\label{IV}}
\subfigure[]{\includegraphics[scale=0.27]{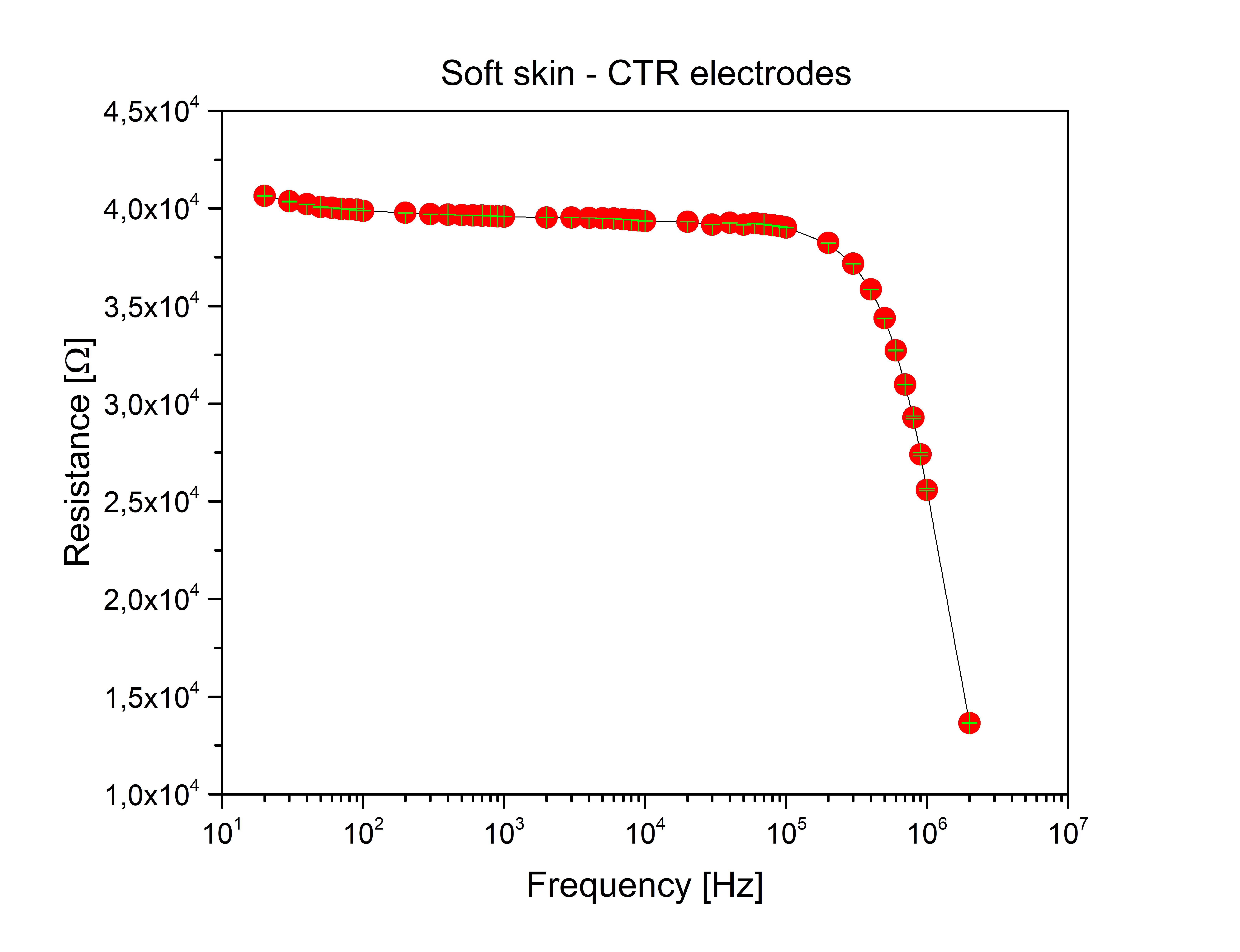}\label{ZRe}}
\subfigure[]{\includegraphics[scale=0.27]{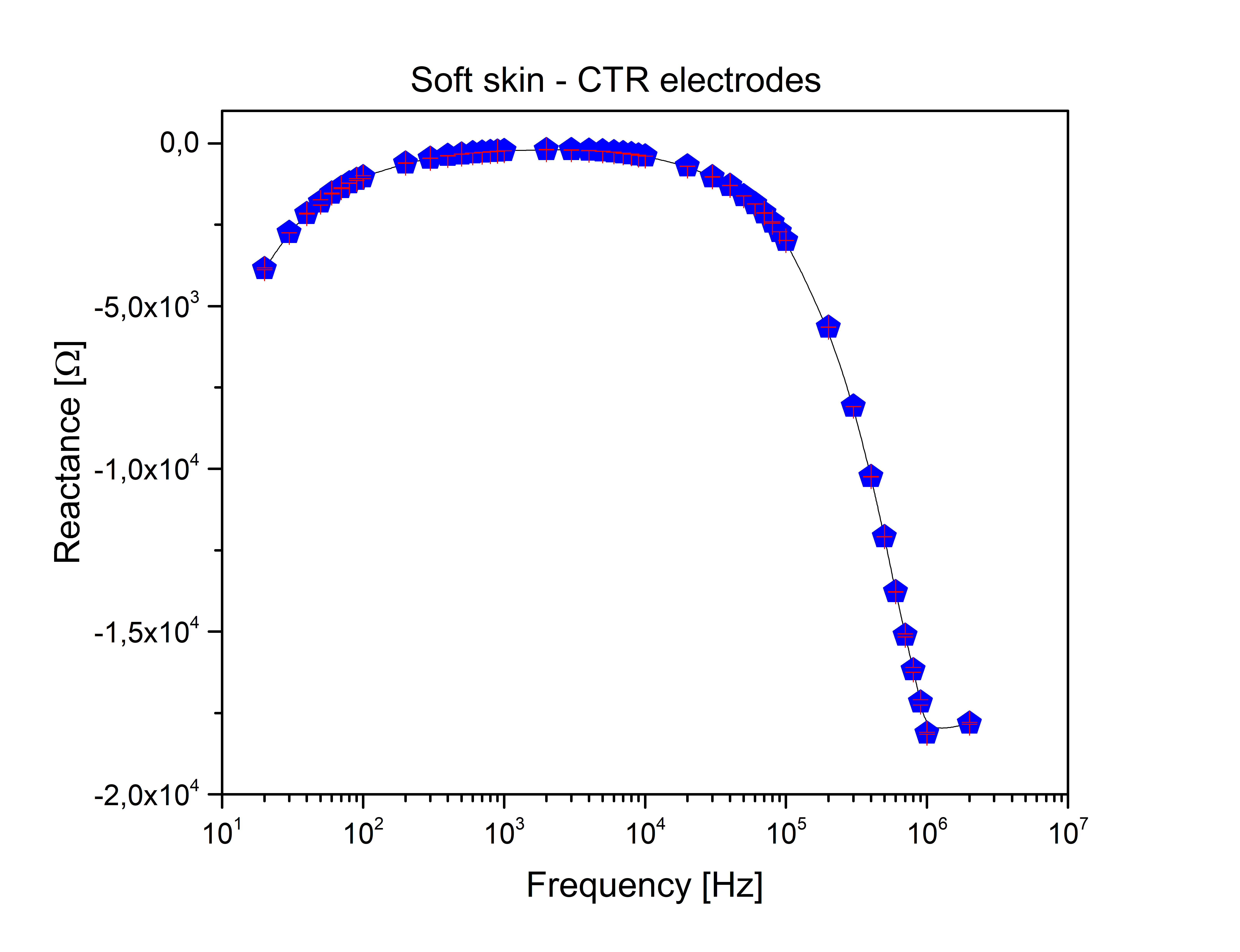}\label{ZIm}}
\caption{The IV curve (left) and AC impedance measured on conductive liquid at rest, hosted in the random network channel of the soft skin.}
\label{fgr:SSR1}
\end{figure}

Two PDMS skins were prepared using different polymers: R10PRO having Shore hardness 10 (soft) and R30PRO having Shore hardness 30 (hard).
For each skin only two electrodes were connected through the upper layer in contact with PEDOT:PSS filled channels: 1) center and bottom left (CBL), 2) top right and bottom left (TRBL). We have found that the choice of electrodes is not fundamental in preserving random network properties: 1) spatio-temporal resolved pressure sensing capabilities and 2) computational capabilities. The IV and impedance characteristic measured show that the system can be modelled as a LCR circuit (see Fig.~\ref{fgr:SSR1}).

\begin{figure}[!tbp]
\centering
  \includegraphics [height=5cm] {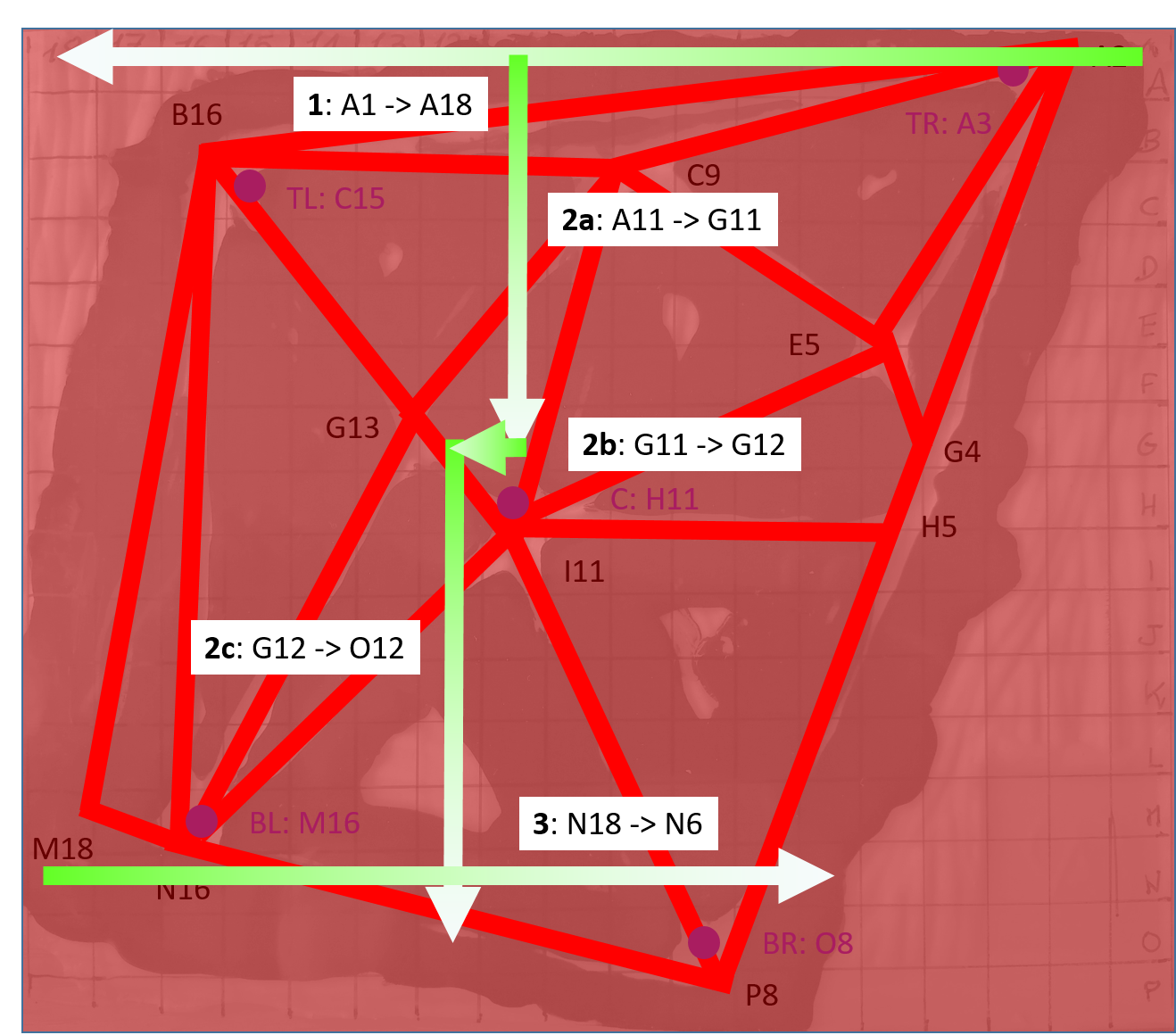}
  \caption{Sketch showing the sequence of boxes submitted to pressure stimuli.}
  \label{fgr:dyna}
\end{figure}

In order to evaluate the spatio-temporal resolution of sensing skin, applying over each box in the chessboard a fixed weight of 100 g was repeated, lifting the weight to move to the next slot after 5 s. During this procedure, impedance was acquired at a fixed frequency by submitting a sinusoidal signal of fixed amplitude of 100~mV, using electrodes in a fixed position. The sequence of boxes where pressure was applied is shown in Fig.~\ref{fgr:dyna}.

\begin{figure}[!tbp]
\subfigure[]{\includegraphics[scale=0.27]{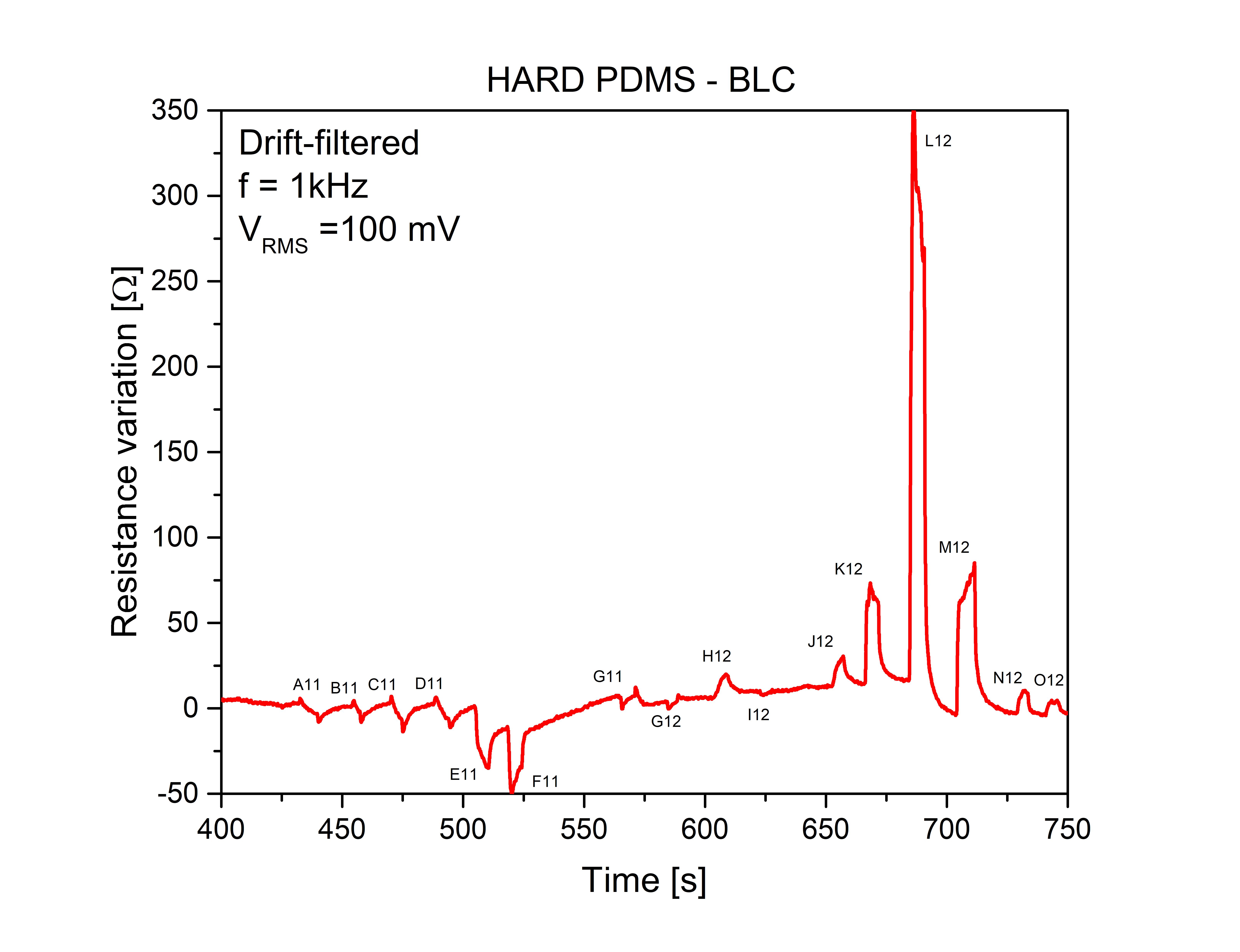}\label{dynaHRe}}
\subfigure[]{\includegraphics[scale=0.27]{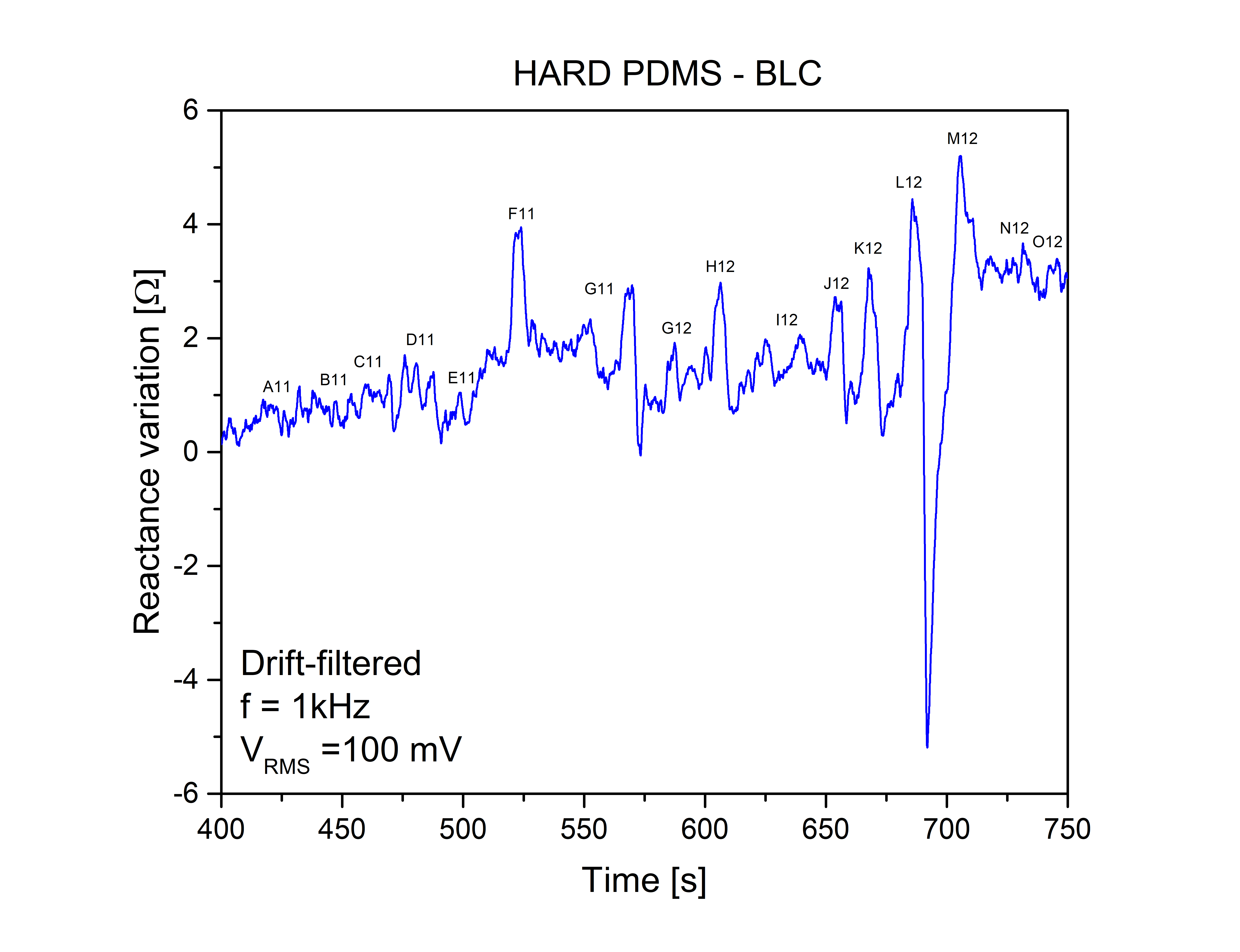}\label{dynaHIm}}
\subfigure[]{\includegraphics[scale=0.27]{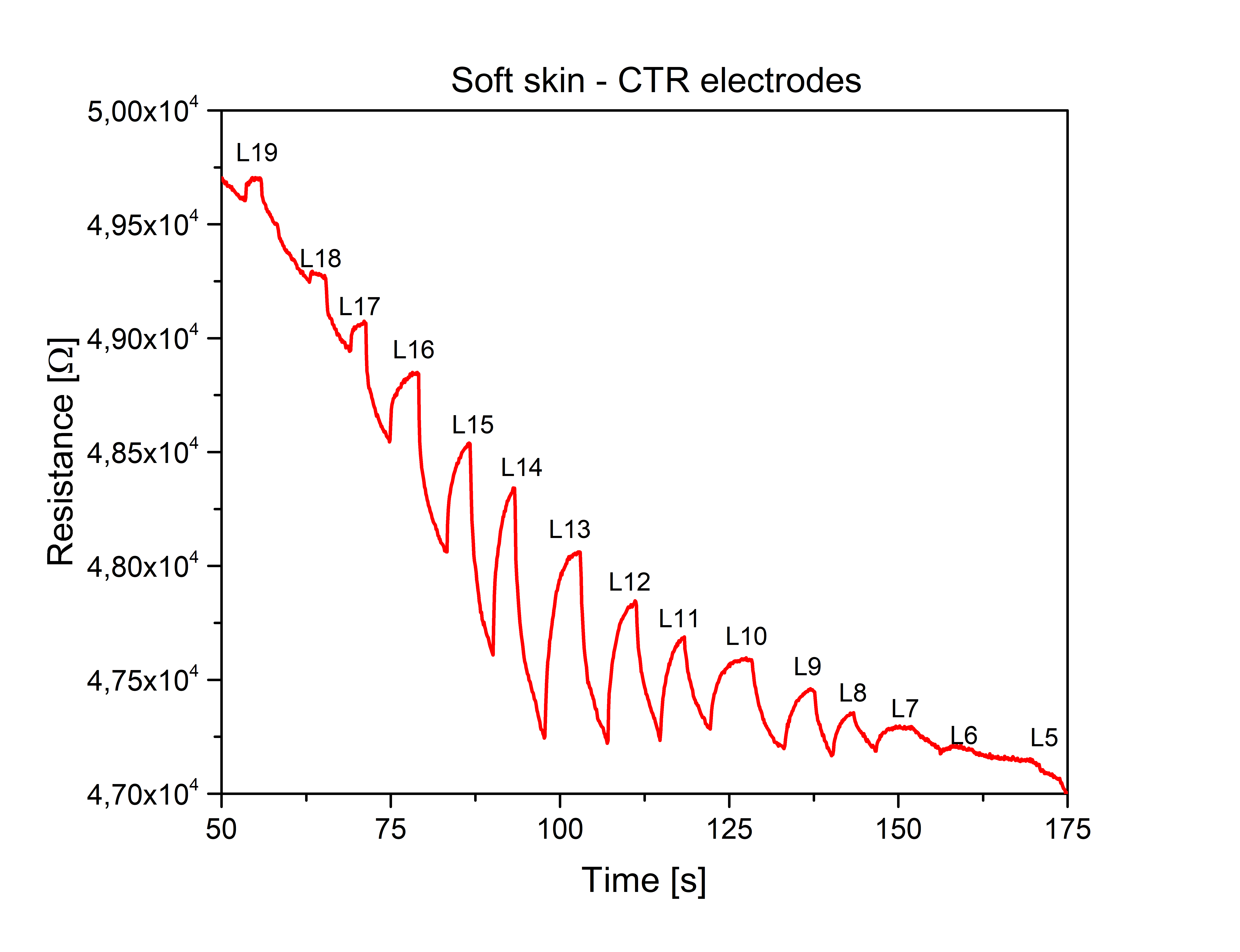}\label{dynaSRe}}
\subfigure[]{\includegraphics[scale=0.27]{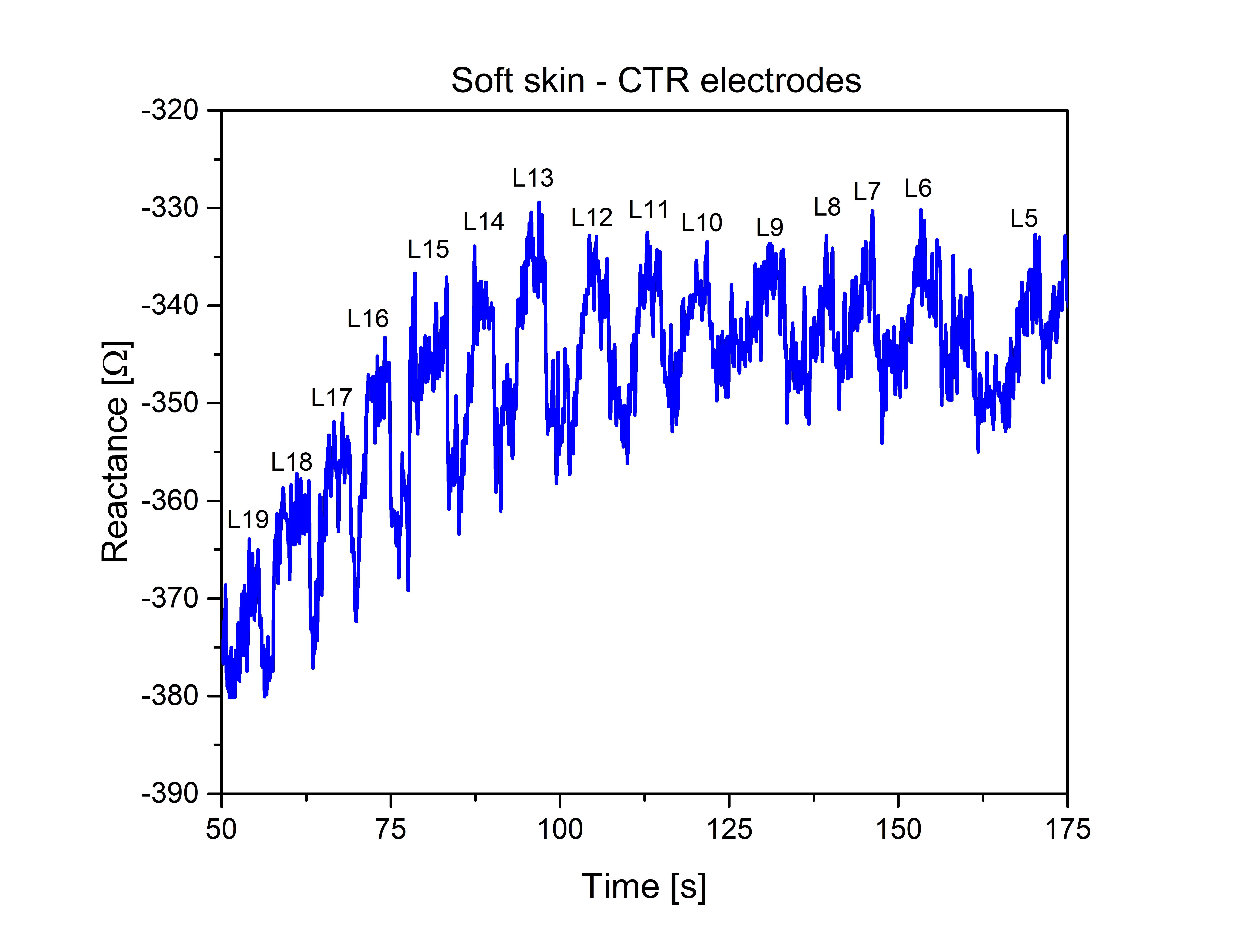}\label{dynaSIm}}
\caption{Top row: pressure stimuli applied to hard skin along path from A11 to O12 passing around center contact point, resistance (a) and reactance (b) collected through bottom left (BL) and center (C) electrodes. Pressure stimuli applied to soft skin along path from L19 to L5 passing around center contact point, resistance (c) and reactance (d) collected through C and top right (TR) electrodes. Both measurements were taken under a sinusoidal signal of 100 mV RMS at a center frequency of 1~kHz and drift during measurements was subtracted algebraically.}
\label{fgr:dyna1}
\end{figure}

\begin{figure}[!tbp]
\centering
  \includegraphics [height=6cm]{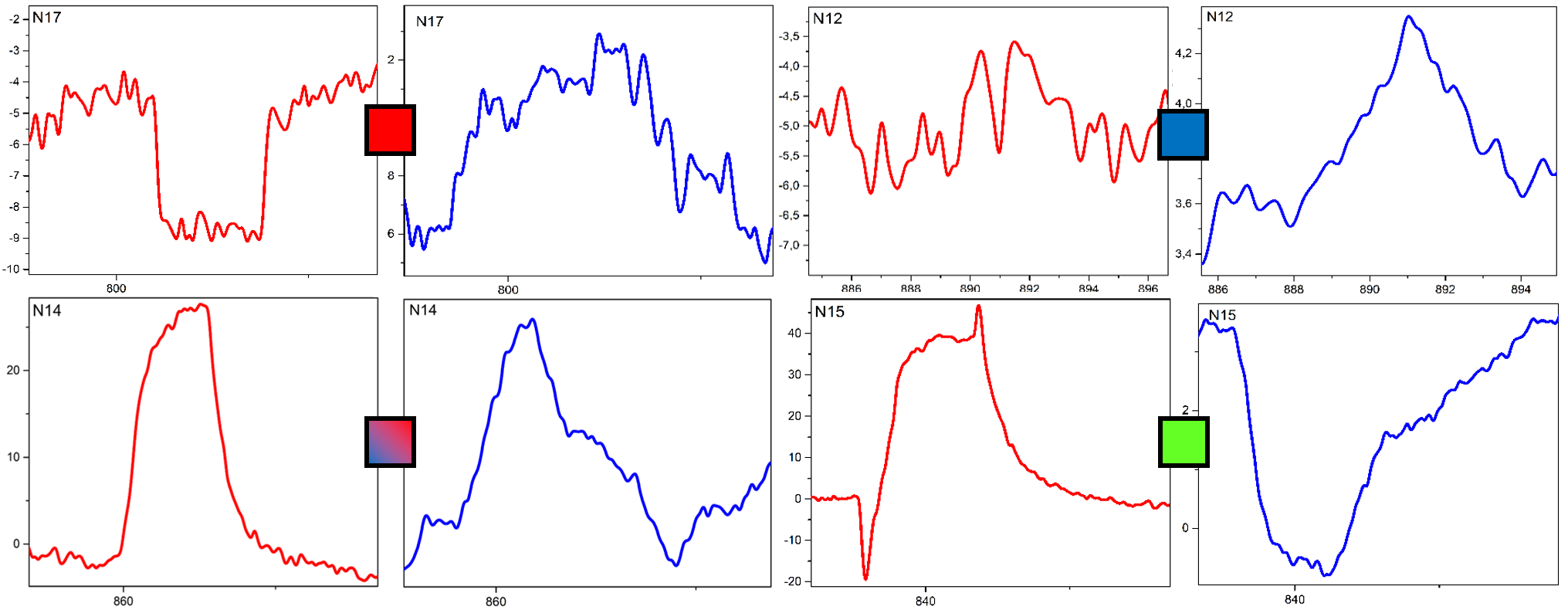}
  \caption{Taxonomy of impedance outputs in response to pressure stimuli, divided into four families. Measurements taken on hard skin with a signal of 100 mV RMS and 1 kHz center frequency.}
  \label{fgr:families}
\end{figure}

The output generated by pressure applied in each position of the chessboard is clearly distinguishable, both reading resistance and reactance (real and imaginary components). All peaks have been labelled as shown in the Fig.~\ref{fgr:dyna1}. Therefore spatial resolution of 1~cm and temporal resolution in the order of 1~s were achieved. By looking at the details of the electronic response to the stimuli (Fig.~\ref{fgr:families}) we can sort the curves into four families (red curve is a resistance, blue curve is a reactance). The first family (RED square) shows a dip in resistance as consequence of applied pressure (while the reactance might show either a peak or a more complex shape); the second family (blue square) shows a peak in the reactance as consequence of applied pressure; the third family (gradient square) belongs to the previous group but shows also a marked peak in resistance; the fourth and last family (green square) shows a dip in reactance (and no features in resistance).

By plotting such responses against the map of the Delaunay triangulation and the electrode location, we see a geometrical correspondence (Fig.~\ref{fgr:dyna}). RED output occurs in boxes which are between electrodes’ projections, meaning that applied pressure creates a deformation able to slightly reduce the distance covered by PEDOT:PSS between electrodes, therefore provoking a major effect on the real part of impedance (resistance). Green output occurs where slots intercept a duct and therefore a pumping effect is possible. External pressure initiates PEDOT:PSS movement and volumetric rearrangements inside the circuit whose effect is that of reducing reactance with almost no changes in resistance. The system can be conceived as a parallel RLC circuit, where around 100~Hz capacitive component prevails, and around 1~MHz inductive component prevails (see Fig.~\ref{fgr:SSR1}). Therefore our stimulus either at 1 or 5~kHz is such that both capacitive and inductive effects are measured. Pumping through applied pressure the fluid in the channels increases the inductive behaviour, like expanding the radiating area of an antenna. It does not happen while crossing channel G13-I11, which is a very short one connecting two major central hubs, and channel I11-M16, which is under the influence of the central electrode. GRADIENT output occurs when the stimulus is applied in between the two actively measuring electrodes: the pressure squeezes the conductive channel, reducing eventually the section of the duct and creating a layered structure where the two polymers are closer, increasing associated capacitance. Blue output occurs in every other situation, meaning that the leading effect is a contraction of the conductive area and a consequent reduction of inductance.

\begin{figure}[!tbp]
\centering
  \includegraphics [height=6cm]{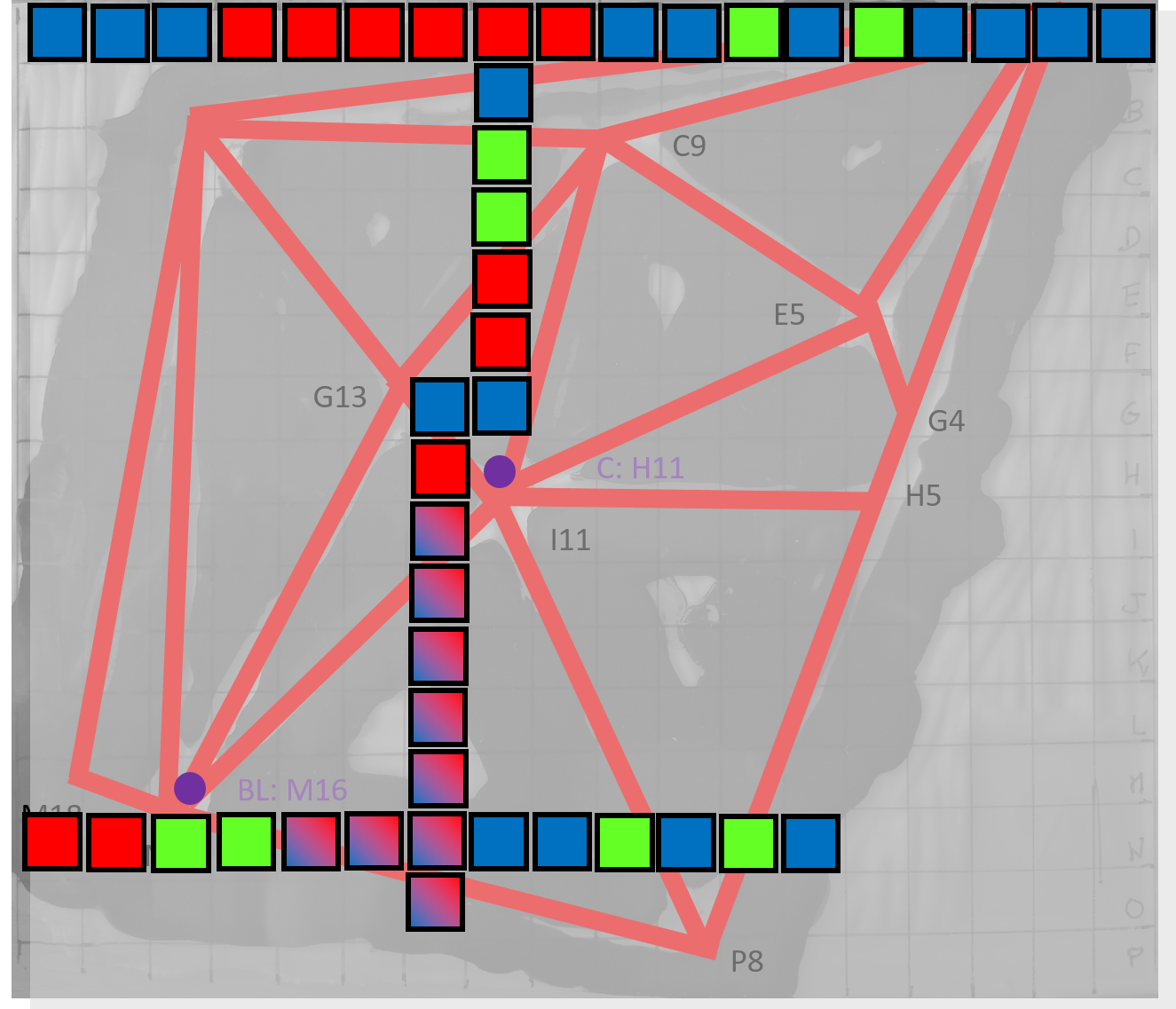}
  \caption{Correspondence map showing the taxonomy of electronic outputs correlated with spatial position of boxes, data collected from the hard skin.}
  \label{fgr:families2}
\end{figure}

\section{Multi-touch experiments for implementing computation}
\label{computation}

\begin{figure}[!tbp]
\centering
  \includegraphics [height=6cm]  {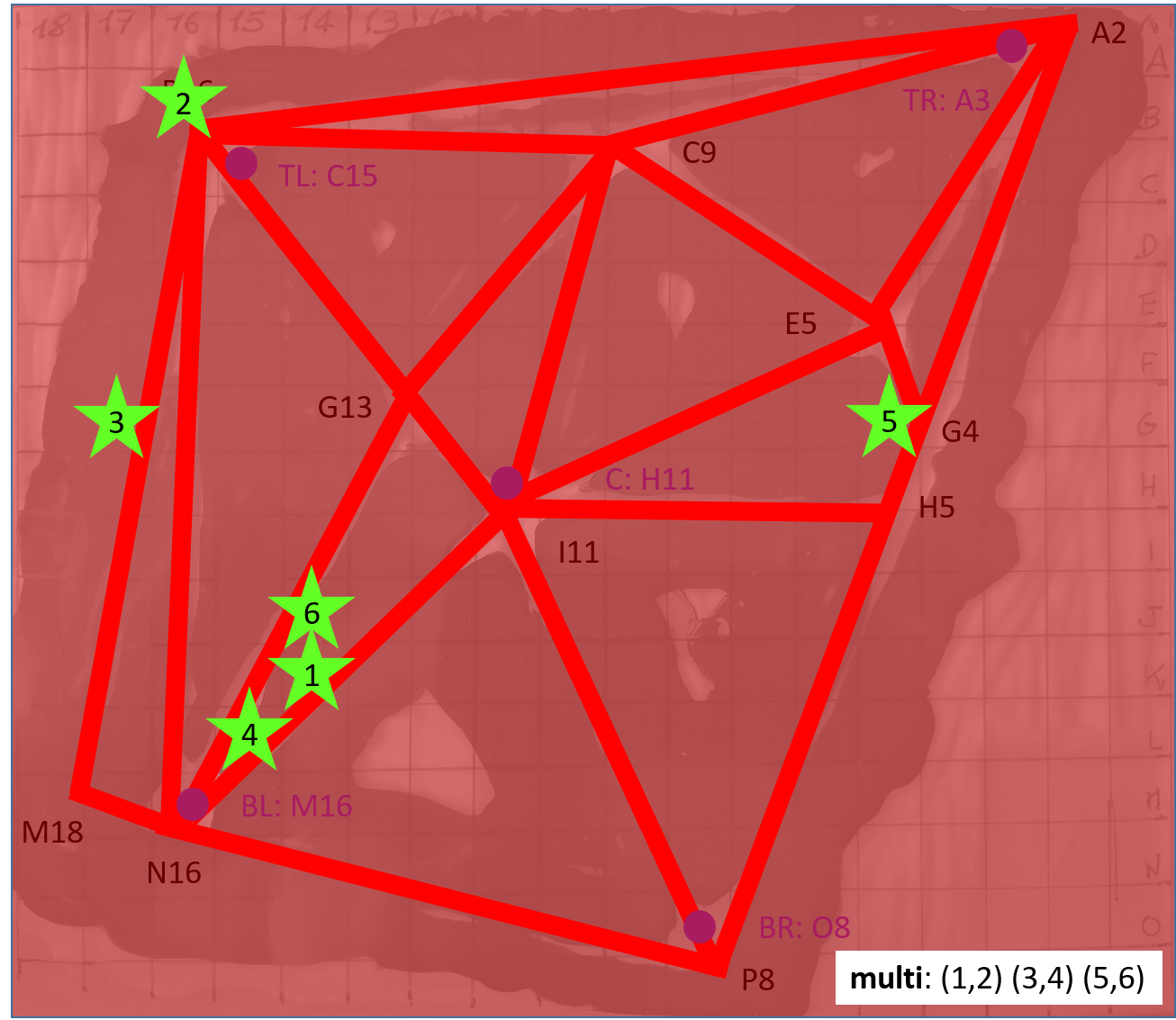}
  \caption{Correspondence map showing the position of boxes and temporal sequential number where pressure stimuli were applied to the skins.}
  \label{fgr:multimap}
\end{figure}

\begin{figure*}[!tbp]
\centering
\subfigure{\includegraphics[height=4cm]{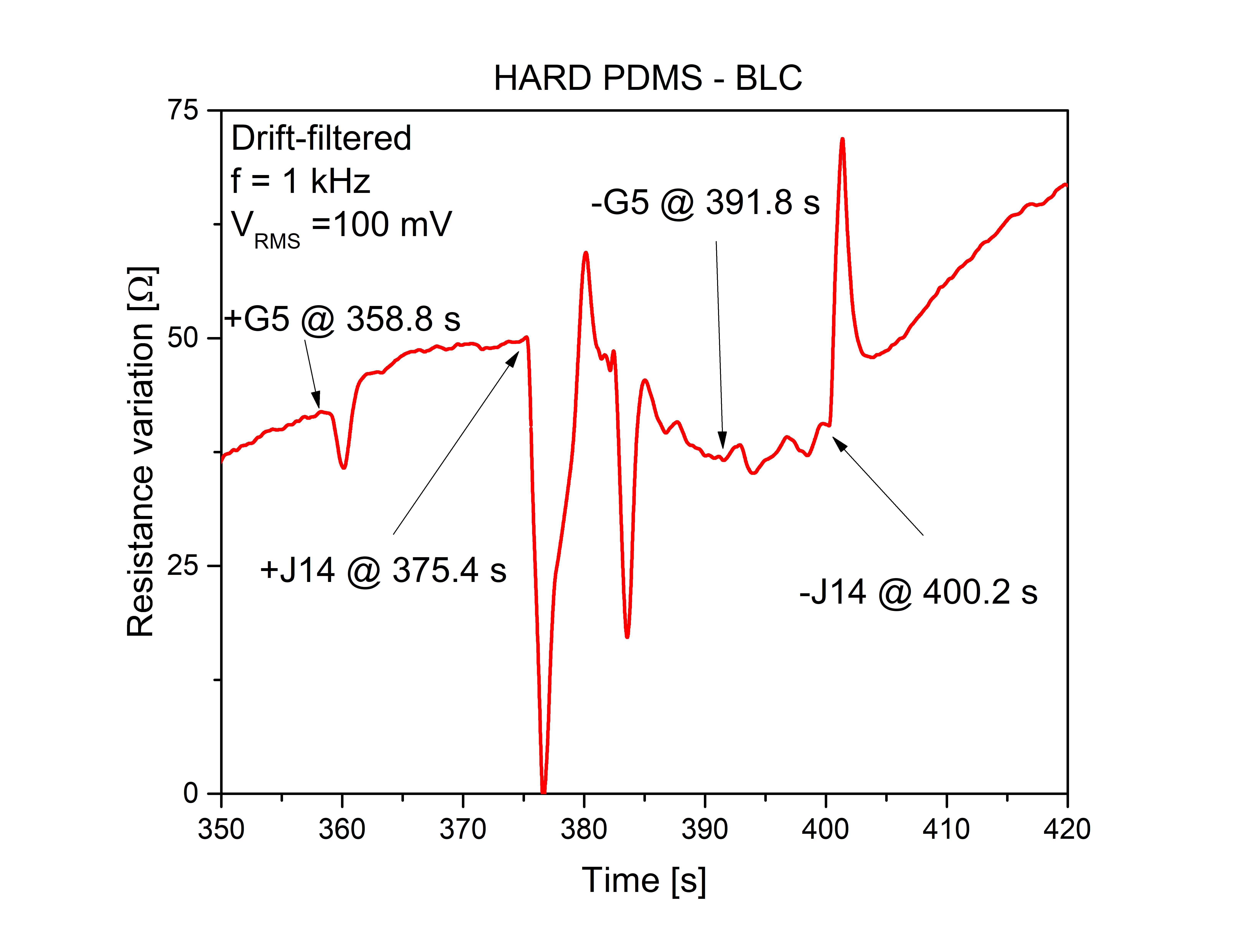}\label{multiRe}}
\subfigure{\includegraphics[height=4cm]{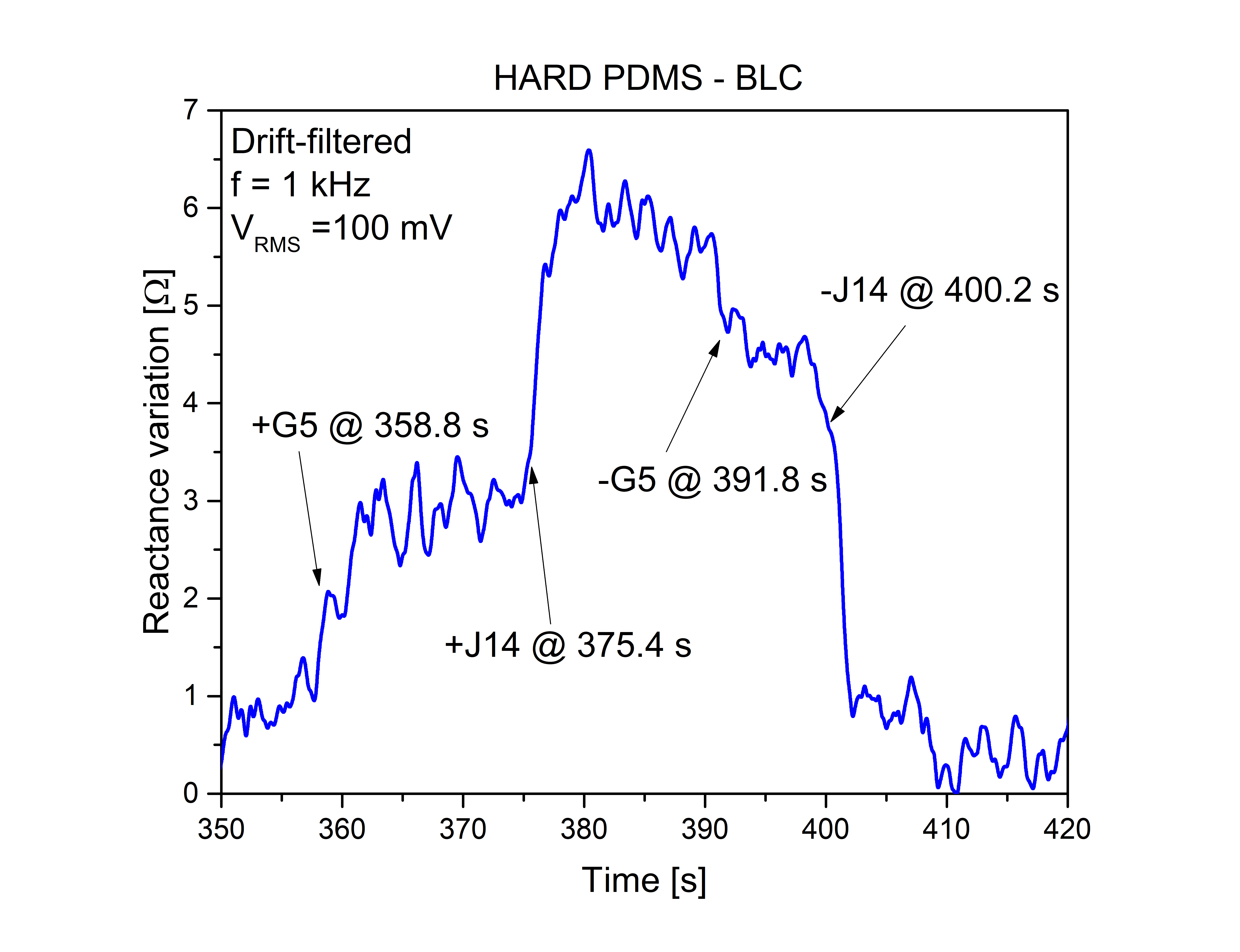}\label{multiIm}}\\
\subfigure{\includegraphics[height=4cm]{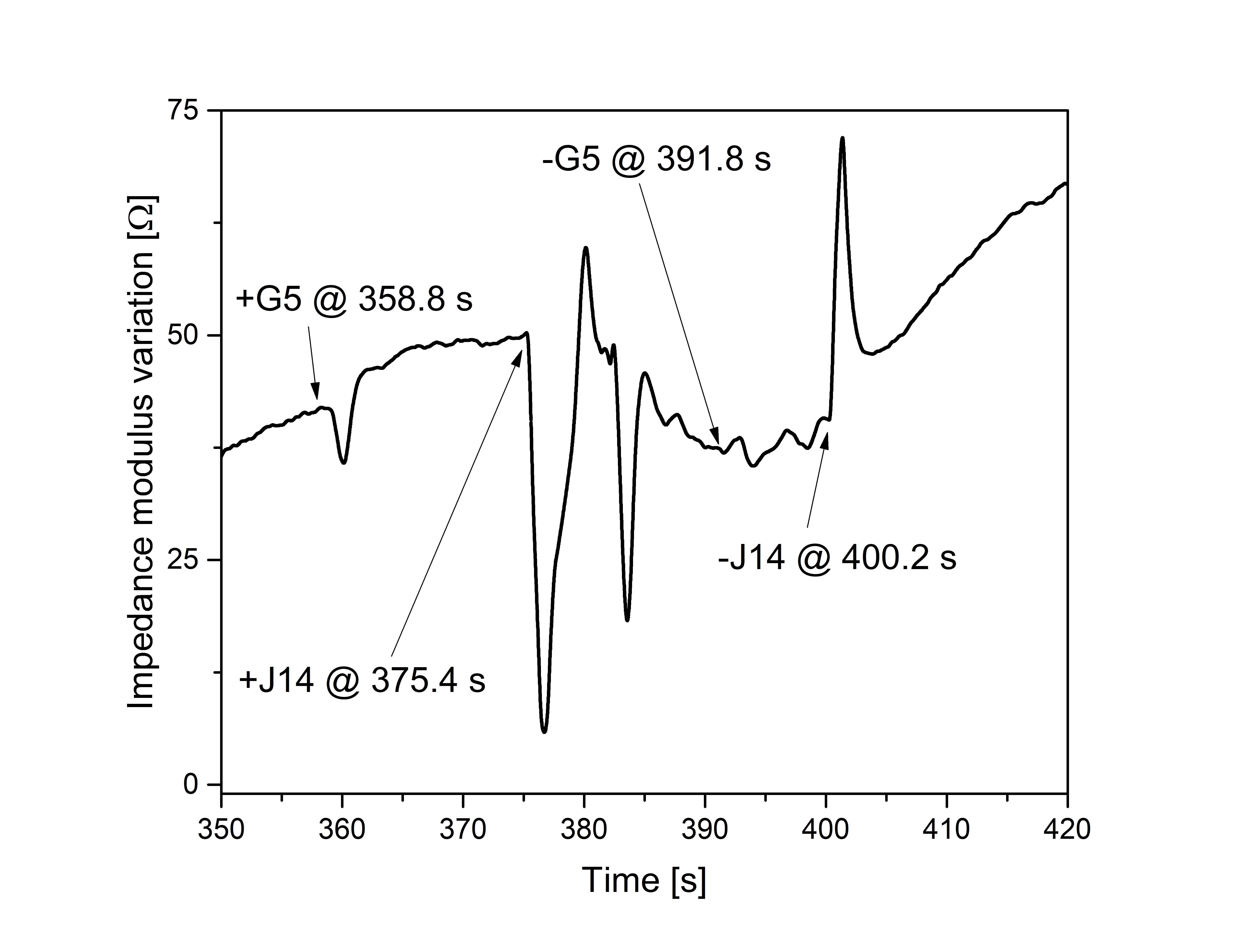}\label{multiMod}}
\subfigure{\includegraphics[height=4cm]{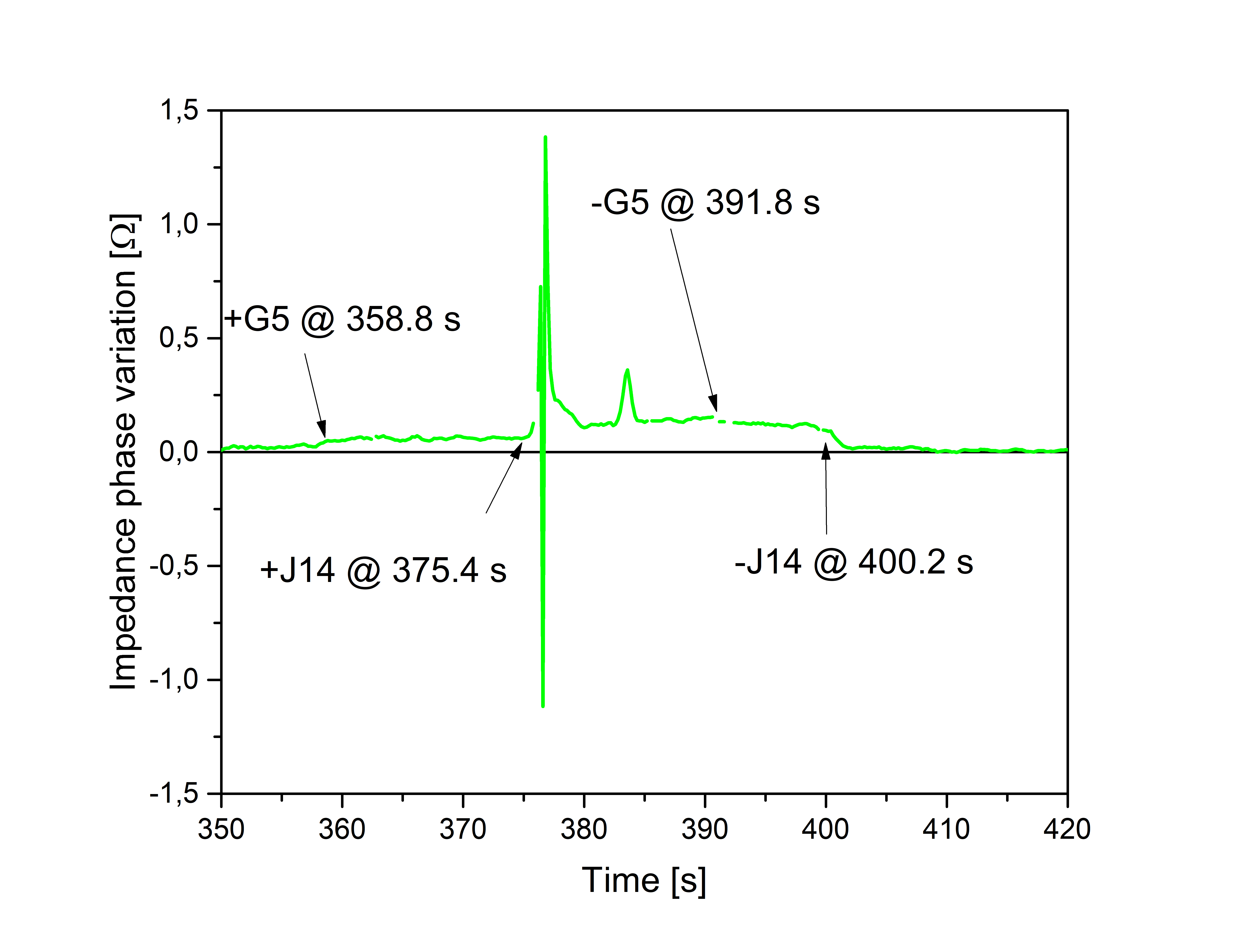}\label{multiPh}}
\caption{From top left to bottom right: resistance, reactance, impedance modulus and impedance phase output measured from multi-touch experiments performed on the hard skin. Measurements taken at 1~kHz, 100~mV RMS, filtered from drifts by algebraic correction.}
\label{fgr:multiout}
\end{figure*}

In the second experiment we implemented a multi-touch protocol: three couples of boxes were identified, regardless of skin hardness and of the positioning of electrodes. For each couple, two pressure stimuli were applied:
\begin{enumerate}
    \item pressure on slot number 1 kept until phase 3;
    \item pressure on slot number 2 kept until phase 4;
    \item relieving pressure from slot number 1;
    \item relieving pressure from slot number 2.
\end{enumerate}
 Electrical output was recorded in real-time during all this process, with a temporal resolution of 200 ms. Figure~\ref{fgr:multimap} shows the location of selected stimuli boxes. The identification of each stimulus can be based on reactance in most of the cases. Also resistance, impedance modulus and impedance phase are shown for completeness in Fig.~\ref{fgr:multiout}.

Here we show how it is possible to associate four different output levels to the four combinations of two inputs:
\begin{itemize}
\item 00: no stimuli are applied, both before and after experiment;
\item 01: first stimulus is removed, second is kept;
\item 10: first stimulus is applied;
\item 11: the two stimuli are applied.
\end{itemize}
The differential reactance, measured on electrodes BL and C, shows that the inputs are separable. The following values of the reactance are recorded:
input $(xy)=(00)$ output $O_{00}=-1.03 \pm 0.05 \Omega$,
input $(xy)=(01)$ output $O_{01}=+5.79 \pm 0.04 \Omega$, 
input $(xy)=(10)$ output $O_{10}=+0.13 \pm 0.03 \Omega$,
input $(xy)=(11)$ output $O_{11}=+8.03 \pm 0.04 \Omega$.
To convert the reactance outputs to binary values, 
$x \times y \xrightarrow{f} z$, where $x,y,z \in \{0, 1\}$ we can impose a cutoff threshold $T$. If $O_{xy}>T$ then $f(x,y)=1$ otherwise $f(x,y)=0$. Thus, for $T=0.13$ we have $f(x,y)=y$ and for $T=5.79$ we have $f(x,y)=xy$.

\section{Discussion}

We demonstrated that a network of channels hosted by a soft polymeric matrix, being a proximity graph constructed on a random planar set, filled with conductive polymer can act as a precise and reliable reporter of the tactile stimuli applied. We have also shown that it is possible to implement Boolean logical circuits, where inputs are represented by positions and strength of mechanic stimuli. Indeed, to be a proper computing the network of channels with conducting polymers must be such that both inputs and outputs share the same physical nature. Nevertheless, at this preliminary stage of research it might be enough that we have achieved computational transduction from mechanical stimuli to electrical outputs. Further studies can go into the direction of complete fusion between sensing and computing, where a stretchable and flexible network of conductive polymers make decision about its response to a complex patterns of dynamic stimuli.

\bibliographystyle{unsrt}
\bibliography{biblio,referencegallium}

\end{document}